\documentclass[sigconf]{acmart}

\renewcommand\footnotetextcopyrightpermission[1]{} % removes footnote with conference information in first column
\usepackage{listings}
\usepackage{courier}
\usepackage{algorithm}
\usepackage{xcolor}
\graphicspath{{figs/}}

\def\BibTeX{{\rm B\kern-.05em{\sc i\kern-.025em b}\kern-.08emT\kern-.1667em\lower.7ex\hbox{E}\kern-.125emX}}
    
\copyrightyear{2019}
\acmYear{2019}
\setcopyright{none}
\acmConference[-]{-}{-}{-}
\acmPrice{15.00}
\begin{document}

%
% The "title" command has an optional parameter, allowing the author to define a "short title" to be used in page headers.
\title{Blaze: Simplified High Performance Cluster Computing}

%
% The "author" command and its associated commands are used to define the authors and their affiliations.
% Of note is the shared affiliation of the first two authors, and the "authornote" and "authornotemark" commands
% used to denote shared contribution to the research.
\author{Junhao Li}
% \authornote{Both authors contributed equally to this research.}
% \orcid{1234-5678-9012}
% \author{G.K.M. Tobin}
% \authornotemark[1]
% \email{webmaster@marysville-ohio.com}
\affiliation{%
  \institution{Cornell University}
%   \streetaddress{P.O. Box 1212}
  \city{Ithaca}
  \state{New York}
  \country{USA}
  \postcode{14853}
}
\email{jl2922@cornell.edu}

\author{Hang Zhang}
\affiliation{%
  \institution{Cornell University}
%   \streetaddress{P.O. Box 1212}
  \city{Ithaca}
  \state{New York}
  \country{USA}
  \postcode{14853}
}
\email{hz459@cornell.edu}

%
% By default, the full list of authors will be used in the page headers. Often, this list is too long, and will overlap
% other information printed in the page headers. This command allows the author to define a more concise list
% of authors' names for this purpose.
\renewcommand{\shortauthors}{Li and Zhang}

%
% The abstract is a short summary of the work to be presented in the article.
% \begin{abstract}
% A clear and well-documented \LaTeX\ document is presented as an article formatted for publication by ACM in 
% a conference proceedings or journal publication. Based on the ``acmart'' document class, this article presents
% and explains many of the common variations, as well as many of the formatting elements
% an author may use in the preparation of the documentation of their work.
% \end{abstract}

\begin{abstract}

MapReduce and its variants have significantly simplified and accelerated the process of developing parallel programs.
However, most MapReduce implementations focus on data-intensive tasks while many real-world tasks are compute intensive and their data can fit distributedly into the memory.
For these tasks, the speed of MapReduce programs can be much slower than those hand-optimized ones.
We present Blaze, a C++ library that makes it easy to develop high performance parallel programs for such compute intensive tasks.
At the core of Blaze is a highly-optimized in-memory MapReduce function, which has three main improvements over conventional MapReduce implementations:
eager reduction, fast serialization, and special treatment for a small fixed key range.
We also offer additional conveniences that make developing parallel programs similar to developing serial programs.
These improvements make Blaze an easy-to-use cluster computing library that approaches the speed of hand-optimized parallel code.
We apply Blaze to some common data mining tasks, including word frequency count, PageRank, k-means, expectation maximization (Gaussian mixture model), and k-nearest neighbors.
Blaze outperforms Apache Spark by more than 10 times on average for these tasks, and the speed of Blaze scales almost linearly with the number of nodes.
In addition, Blaze uses only the MapReduce function and 3 utility functions in its implementation while Spark uses almost 30 different parallel primitives in its official implementation.

\end{abstract}

%
% The code below is generated by the tool at http://dl.acm.org/ccs.cfm.
% Please copy and paste the code instead of the example below.
% %
\begin{CCSXML}
<ccs2012>
<concept>
<concept_id>10010147.10010169.10010170.10003817</concept_id>
<concept_desc>Computing methodologies~MapReduce algorithms</concept_desc>
<concept_significance>500</concept_significance>
</concept>
<concept>
<concept_id>10002951.10003227.10003351</concept_id>
<concept_desc>Information systems~Data mining</concept_desc>
<concept_significance>300</concept_significance>
</concept>
<concept>
<concept_id>10002951.10003227.10003351.10003444</concept_id>
<concept_desc>Information systems~Clustering</concept_desc>
<concept_significance>300</concept_significance>
</concept>
<concept>
<concept_id>10002951.10003227.10003351.10003445</concept_id>
<concept_desc>Information systems~Nearest-neighbor search</concept_desc>
<concept_significance>300</concept_significance>
</concept>
<concept>
<concept_id>10002951.10003260.10003261.10003263.10003265</concept_id>
<concept_desc>Information systems~Page and site ranking</concept_desc>
<concept_significance>300</concept_significance>
</concept>
<concept>
<concept_id>10010405.10010432</concept_id>
<concept_desc>Applied computing~Physical sciences and engineering</concept_desc>
<concept_significance>100</concept_significance>
</concept>
<concept>
<concept_id>10010405.10010432.10010441</concept_id>
<concept_desc>Applied computing~Physics</concept_desc>
<concept_significance>100</concept_significance>
</concept>
</ccs2012>
\end{CCSXML}

\ccsdesc[500]{Computing methodologies~MapReduce algorithms}
\ccsdesc[300]{Information systems~Data mining}
\ccsdesc[300]{Information systems~Clustering}
\ccsdesc[300]{Information systems~Nearest-neighbor search}
\ccsdesc[300]{Information systems~Page and site ranking}
% \ccsdesc[100]{Applied computing~Physical sciences and engineering}
% \ccsdesc[100]{Applied computing~Physics}

%
% Keywords. The author(s) should pick words that accurately describe the work being
% presented. Separate the keywords with commas.
\keywords{MapReduce, high performance, cluster computing, data mining, PageRank, k-means, expectation maximization, Gaussian mixture, k-nearest neighbors, serialization}

%
% A "teaser" image appears between the author and affiliation information and the body 
% of the document, and typically spans the page. 
% \begin{teaserfigure}
%   \includegraphics[width=\textwidth]{sampleteaser}
%   \caption{Seattle Mariners at Spring Training, 2010.}
%   \Description{Enjoying the baseball game from the third-base seats. Ichiro Suzuki preparing to bat.}
%   \label{fig:teaser}
% \end{teaserfigure}

%
% This command processes the author and affiliation and title information and builds
% the first part of the formatted document.
\maketitle

\section{Introduction}

Cluster computing enables us to perform a huge amount of computations on big data and get insights from them at a scale that a single machine can hardly achieve.
However, developing parallel programs to take advantage of a large cluster can be very difficult.

\begin{figure}
  \begin{center}
  \includegraphics[width=\linewidth]{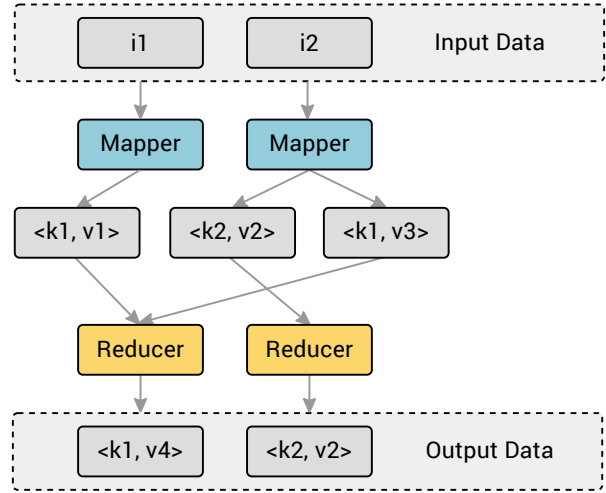}
  \end{center}
  \vspace{-0.2cm}
  \caption{MapReduce Programming Model.
  The map function generates a set of intermediate key/value pairs for each input.
  The reduce function merges the values associated with the same key.
  Numerous data mining and machine learning algorithms are expressible with this model.
  }
  \label{fig:mr}
\end{figure}

MapReduce~\cite{dean2008mapreduce,dean2010mapreduce} greatly simplified this task by providing users a high-level abstraction for defining their computation, and taking care of the intricate low-level execution steps internally.
Fig.~\ref{fig:mr} illustrates the MapReduce programming model.
Logically, each MapReduce operation consists of two phases: a map phase where each input is mapped to a set of intermediate key/value pairs, and a reduce phase where the pairs with the same key are put together and reduced to a single key/value pair according to a user specified reduce function.

Many data mining algorithms are expressible with this model, such as PageRank~\cite{bahmani2011fast,plimpton2011mapreduce,ekanayake2010twister}, k-means~\cite{zhao2009parallel,chu2007map,cui2014optimized,anchalia2013mapreduce,ekanayake2008mapreduce,gopalani2015comparing}, Gaussian mixture model~\cite{chu2007map}, and k-nearest neighbors~\cite{anchalia2014k,maillo2015mapreduce,lu2012efficient,yokoyama2012processing}.

Although logically expressible, achieving similar efficiency as a hand-optimized parallel code is hard, especially when the data can be fit distributed into the memory.
In such cases, the file system is no longer the bottleneck and the overhead from MapReduce can make the execution much slower than hand-optimized code.

Google's MapReduce~\cite{dean2008mapreduce,dean2010mapreduce} and most of its variants~\cite{hadoop,chambers2010flumejava,cascading,afrati2010optimizing,bhatotia2011incoop,condie2010mapreduce,ekanayake2010twister,goiri2015approxhadoop,he2008mars,li2015coded,zaharia2008improving,yang2007map} save intermediate data and result to the file system even when the data can be fit into the memory.
Hence, its MapReduce performance is severely limited by the performance of the file system.

Spark~\cite{spark, zaharia2010spark, zaharia2016apache, zaharia2012resilient} offers an in-memory implementation of MapReduce, which is much faster than Google's MapReduce.
However, it uses a similar algorithm as Google's MapReduce, which is designed for disk-based data intensive use cases and does not consider the computational overheads of MapReduce seriously.
Hence, the performance of Spark is often far from the performance of hand-optimized code.

To achieve better performance while preserving the high-level MapReduce abstraction, we develop Blaze, a C++ based cluster computing library that focuses on in-memory high performance MapReduce and related operations.
Blaze introduces three main improvements to the MapReduce algorithm: eager reduction, fast serialization, and special treatment for a small fixed key range.
Section~\ref{sec:opt} provides a detailed description of these improvements.

% Blaze is  quantum chemistry package~\cite{li2018fast}.
% Blaze has already been used in a quantum chemistry package~\cite{li2018fast} to make the code more maintainable without sacrificing the performance.
% Together with that quantum chemistry package, Blaze has been run for tens of millions of core-hours by several top research groups to perform highly accurate quantum simulations and aid the discovery of new materials and new chemical processes.

% The highly-optimized in-memory MapReduce implementation can be very useful to numerous data mining and machine learning related tasks.

We apply Blaze to several common data mining tasks, including word frequency count, PageRank, k-means, expectation maximization (Gaussian mixture), and k-nearest neighbors.
Our results show that Blaze is on average 10 times faster than Spark on these tasks.
% We compare its performance with Apache Spark and results

The main contributions of this research are listed as follows:
\begin{enumerate}
    \item We develop Blaze, a high performance cluster computing library that allows users to write parallel programs with the high-level MapReduce abstraction while achieving similar performance as hand-optimized code for compute intensive tasks.
    \item We introduce three main performance improvements to the MapReduce algorithm to make it more efficient: eager reduction, fast serialization, and special treatment for a small fixed key range.
    \item We apply Blaze to 5 common data mining tasks and demonstrate that Blaze programs are easy to develop and can outperform Apache Spark programs by more than 10 times on average for these tasks.
    % by applying the improvements above, we can keep the simple high-level abstraction of MapReduce while achieving $10 \times$ higher performance than Apache Spark on some common data mining and machine learning tasks.
\end{enumerate}

The remaining sections are organized as follows:
Section~\ref{sec:blaze} describes the Blaze framework and the details of the optimization.
Section~\ref{sec:app} present the details of how we implement several key data mining and machine learning algorithms with Blaze and compare the performance with Apache Spark.
Section~\ref{sec:con} concludes the paper.

\section{The Blaze Library}
\label{sec:blaze}
The Blaze library offers three sets of APIs: 1) a high-performance MapReduce function, 2) distributed data containers, and 3) parallel computing utility functions.
These APIs are built based on the Blaze parallel computing kernel, which provides common low-level parallel computing primitives.

% Blaze also offers several utility functions for parallel programming, including thread-safe random number generators, a parallel file loader, and converters between C++ standard library containers and Blaze distributed data containers.

\begin{figure}
  \begin{center}
  \includegraphics[width=\linewidth]{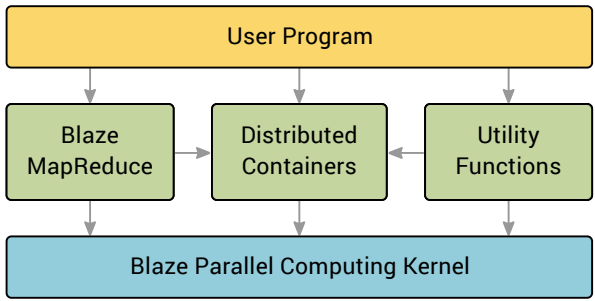}
  \end{center}
  \vspace{-0.2cm}
  \caption{Blaze Architecture.
%   Blaze exposes three sets of APIs: the high performance Blaze MapReduce function, distributed data containers, and a few utility functions.
%   The Blaze parallel computing kernel provides low-level parallel computing primitives to support the high-level APIs.
  }
  \label{fig:mrdiff}
\end{figure}

\subsection{Distributed Containers}

Blaze provides three distributed data containers: \emph{DistRange}, \emph{DistVector}, and \emph{DistHashMap}.
DistRange does not store the whole data but only the start, the end, and the step size of the data.
DistVector distributedly stores an array of elements.
DistHashMap distributedly stores key/value pairs.

All of the three containers support the \lstinline{foreach} operation, where a custom function can be applied to each of its element in parallel.
This function can either change the value of the element itself or use the value of the element to perform external operations.

Both the DistVector and the DistHashMap can be converted to and from C++ standard library containers with Blaze utility functions \lstinline{distribute} and \lstinline{collect}.
DistVector can also be created from the \lstinline{load_file} utility function, which can load text files from the file system parallelly into a distributed vector of lines.

DistVector also has a \lstinline{topk} method, which can return the top k elements from the distributedly stored vector in $O(n+k\log k)$ time and $O(k)$ space.
Users can provide a custom comparison function to determine the priority of the elements.

\subsection{MapReduce}

The MapReduce function uses a functional style interface.
It takes four parameters:
\begin{enumerate}
    \item Input. One of the Blaze distributed container.
    \item Mapper. When the input is a DistRange, the mapper should be a function that accepts two parameters: a value from the DistRange and a handler function for emitting key/value pairs.
    When the input is a DistVector or a DistHashMap, the mapper should be a function that accepts three parameters: a key from the input, the corresponding value, and an emit handler.
    \item Reducer. The function that reduce two values to one value.
    Blaze provides several built-in reducers, including \lstinline{sum}, \lstinline{prod}, \lstinline{min}, and \lstinline{max}, which can cover most use cases.
    These reducers can be used by providing the reducer name as a string, for example, \lstinline{"sum"}.
    Users can also provide custom reduce functions, which should take two parameters, the first one is a reference to the existing value which needs to be updated, and the second one is a constant reference to the new value.
    \item Target. One of the Blaze distributed container or a vector from the standard library.
    The target container should be mutable and it is not cleared before performing MapReduce.
    New results from the MapReduce operation are merged/reduced into the target container.
\end{enumerate}

Blaze MapReduce also takes care of the serialization of common data types so that the map function can emit non-string key/value pairs, and the reduce function no longer requires additional logic for parsing the serialized data.
Using custom data types as keys or values is also supported. For that, users only need to provide the corresponding serialize/parse methods and a hash function (for keys).

We provide two examples of using Blaze MapReduce in Appendix~\ref{app:wordcount} and \ref{app:pi}.

\subsection{Optimization}
\label{sec:opt}
We introduce several optimizations to make the MapReduce function faster, including eager reduction, fast serialization, and special treatment for cases where the resulting key range is small and fixed.

\subsubsection{Eager Reduction}

Conventional MapReduce performs the map phase first and saves all the emitted pair from the mapper function.
Then, it shuffles all the emitted pairs across the networks directly, which could incur a large amount of network traffics.

\begin{figure}
  \begin{center}
  \includegraphics[width=\linewidth]{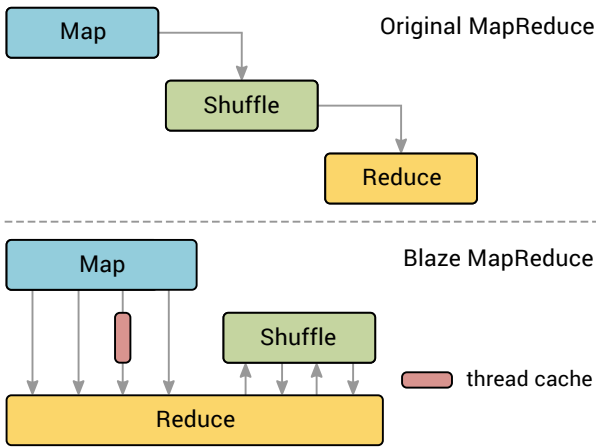}
  \end{center}
  \vspace{-0.2cm}
  \caption{Eager Reduction in Blaze MapReduce.
%   Blaze performs machine-local reduce right after the mapper function emits a key/value pair.
%   For popular keys, Blaze automatically reduce new values to a thread-local cache instead of the main copy on a machine.
%   The cross-machine shuffle operates on the locally reduced data and thus requires much less network communication.
%   During the shuffle operations, reduce operations also happen asynchronously to maximize the throughput.
  }
  \label{fig:mrdiff}
\end{figure}

In Blaze MapReduce, we perform machine-local reduce right after the mapper function emits a key/value pair.
For popular keys, Blaze automatically reduces new values to a thread-local cache instead of the machine-local copy.
The cross-machine shuffle operates on the locally reduced data which substantially reduces the network communication burden.
During the shuffle operations, reduce functions are also operating asynchronously to maximize the throughput.
Fig.~\ref{fig:mrdiff} illustrates the difference between the conventional MapReduce and Blaze MapReduce with eager reduction.

\subsubsection{Fast Serialization}
During the shuffle/reduce phase, we serialize the messages into a compact binary format before casting them across the network.

Our encoding scheme and algorithm are similar to Google's Protobuf~\cite{protobuf} but without prefixing each entry with field tags and wire types.
Although these two fields allow missing fields and support serializing the fields in arbitrary order, this additional flexibility is not needed in MapReduce.
On the other hand, these two fields can have a significant impact on both the performance and the serialized message size, especially when the content size of a field is small, which is common for MapReduce key/value pairs.
For example, when both the key and value are small integers, the serialized message size of each pair from Protocol Buffers will be 4 bytes while the message from Blaze fast serialization will be only 2 bytes, which is 50\% smaller than the one from Protocol Buffers.
Removing the fields tags and wire types does not cause ambiguity as long as we always serialize the fields in the same order, which is easy to achieve in MapReduce.
The smaller size in the serialized message means less network traffics, so that Blaze can scale better on large clusters when the cross-rack bandwidth becomes the bottleneck.

% \subsubsection{Hash-Based Shuffle}

% We use hash based shuffle instead of the original sort based shuffle.
% Hash based shuffle can send the messages to the corresponding worker in $O(1)$ time.

% Note that in the MapReduce results, the keys are no longer sorted.
% We provide a separate function, called \lstinline{top} for achieving similar capability in the distributed vector container.
% \lstinline{top} returns the top k elements in the distributed vector container in $O(n + k \log k)$ time, according to a custom compare function.
% When $k=n$, it will essentially perform a parallel merge sort.
% In section \ref{sec:nn}, we provide an example of the nearest 100 neighbors related to a given point from a huge set of other points, which uses the member function.

\subsubsection{Optimization for Small Key Range}

For small key range, we create a thread-local cache for each key at the beginning and set that as the reduce target during the local map/reduce phase.
After the local map/reduce phase finished, we perform parallel tree based reduce operations: first locally and then across multiple machines.
The resulting execution plan is essentially the same as hand-optimized parallel for loops with thread-local intermediate results.

\begin{table}
  \caption{Monte Carlo Pi Estimation Performance.
  We can see that Blaze MapReduce has almost the same speed as hand-optimized MPI+OpenMP parallel for loops while requires much fewer source lines of code (SLOC).}
  \label{tab:pi}
  \begin{tabular}{ccc}
    \toprule
    Samples & Blaze MapReduce & MPI+OpenMP\\
    \midrule
    $10^7$ & $0.14\pm 0.01$ s& $0.14\pm 0.01$ s \\
    $10^8$ & $1.44\pm 0.07$ s& $1.42\pm 0.09$ s \\
    $10^9$ & $14.2\pm 1.3$ s& $14.6\pm 1.7$ s \\
    \midrule
    SLOC & 8 & 24 \\
  \bottomrule
\end{tabular}
\end{table}

We benchmark the performance of Blaze MapReduce against hand-optimized parallel for-loop on the Monte Carlo Pi estimation task.
In this task, the mapper function first generates two random numbers $x$ and $y$ in the range $[0, 1]$, and then emits 1 to key 0 when $x^2 + y^2 < 1$.
Cases like this where we reduce big data to a small number of keys are commonly seen in data mining and are not efficient with the original MapReduce algorithm.
However, by using a thread-local copy as the default reduce target for each thread, Blaze MapReduce can achieve similar performance as hand-optimized code based on raw MPI and OpenMP.
Table~\ref{tab:pi} reports the result and Appendix~\ref{app:pi} lists our implementation.
The tests are performed on a local machine with Ubuntu 16.04, GCC 5.4 -O3, and an Intel i7-8550U processor.

\section{Applications}
\label{sec:app}
% Blaze is originally designed for quantum simulation to make the code more readable and extensible while perserving the performance of hand-optimized parallel code.
% We achieved this by adapt our quantum simulation algorithm into a series of MapReduce operations and create a highly optimized MapReduce implementation.

% We find that our implentation can also be useful for many data mining tasks as well, because many data mining tasks share the same kind of MapReduce workflows as our quantum simulation algorithm and our implementation can allow us to write more concise and extensible code with the high-level MapReduce API while achieving the performance of hand-optimized parallel C++ code.
In this section, we benchmark Blaze against a popular data mining package Spark, on common data mining tasks, including word frequency count, PageRank, k-means, expectation maximization (with the Gaussian Mixture model), and k-nearest neighbors search.

\subsection{Task Description and Implementation}

In this section, we describe the data mining tasks and how we implement them in Blaze and Spark.
All the source code of our implementation is included in our GitHub repository~\cite{blaze}.

\subsubsection{Word Frequency Count}

This task counts the number of occurrences of each unique English words in a text file.
We use the Bible and Shakespeare's works as the testing text.
Since Spark has significant overhead in starting the MapReduce tasks, we repeat the Bible and the Shakespeare 200 times, so that the input file contains about 0.4 billion words.

We use MapReduce in both Blaze and Spark.
The mapper function takes a single line and emits multiple (word, 1) pairs.
The reducer function sums the values.
Appendix \ref{app:wordcount} contains the full Blaze implementation for this example.

\subsubsection{PageRank}

This task calculates the PageRank score, which is defined as the stationary value of the following equation:
\begin{equation}
    \label{eq:pr}
    PR(p_i) = \frac{1-d}{N} + d \sum_{p_j\in M(p_i)} \frac{PR(p_j)}{L(p_j)}
\end{equation}
where $M(p_i)$ is the set of pages that link to $p_i$, $L(p_j)$ is the number of outbound links from page $p_j$, $N$ is the total number of pages, and $d=0.15$.
When a page has no outbound links, it is called a sink and is assumed to connect to all the pages.
We use the graph500 generator to generate the input graph which contains 10 million links.
We set the convergence criterion to $10^{-5}$, which results in 27 iterations for our input.
The links are stored distributedly across multiple machines.

For Blaze, we use 3 MapReduce operations per iteration to implement this task.
The first one calculates the total score of all the sinks.
The second one calculates the new PageRank scores according to Eq.~\ref{eq:pr}.
The third one calculates the maximum change in the scores of all the pages.
For Spark, we use the built-in PageRank module from the Spark GraphX library~\cite{xin2013graphx}.

\subsubsection{K-Means}

K-Means is a popular clustering algorithm.
The algorithm proceeds by alternating two steps until the convergence.
The first step is the assignment step where each point is assigned to the nearest clustering center.
The second step is the refinement step where each clustering center is updated based on the new mean of the points assigned to the clustering center.

We generate 100 million random points around 5 clustering centers as the testing data, and use the same initial model and convergence criteria for Spark and Blaze.
The points are stored distributedly across multiple machines.

For Blaze, we use a single MapReduce operation to perform the assignment step.
The update step is implemented in serial.
For Spark, we use the built-in implementation from the Spark MLlib library~\cite{meng2016mllib}.

\subsubsection{Expectation Maximization}

This task uses the expectation maximization method to train the Gaussian Mixture clustering model (GMM).
Starting from an initial model, we first calculate the Gaussian probability density of each point for each Gaussian component
\begin{equation}
    \label{eq:prob}
    p _{ k } \left( \vec{ x } | \theta _{ k } \right) = \frac{ 1 }{ \left( 2 \pi \right) ^{ d / 2 } \left| \Sigma _{ k } \right| ^{ 1 / 2 } } e ^{ - \frac{ 1 }{ 2 } \left( \vec{ x } - \vec{ \mu } _{ k } \right) ^{ T } \Sigma _{ k } ^{ - 1 } \left( \vec{ x } - \vec{ \mu } _{ k } \right) }
\end{equation}
where $\mu_1$ to $\mu_K$ are the centers of these Gaussian components and $\Sigma_1$ to $\Sigma_K$ are the covariance matrices.
Then we calculate the membership of each point for each Gaussian component
\begin{equation}
    \label{eq:membership}
    w _{ i k } = \frac{ p _{ k } \left( \vec{ x } _{ i } | \theta _{ k } \right) \cdot \alpha _{ k } }{ \sum _{ m = 1 } ^{ K } p _{ m } \left( \vec{ x } _{ i } | \theta _{ m } \right) \cdot \alpha _{ m } }
\end{equation}
where $\alpha_k$ is the weights of the Gaussian component.
Next, we calculate the sum of membership weights for each Gaussian component $N_k=\Sigma_{i=1}^K w_{ik}$.
After that, we update the parameters of the Gaussian mixtures
\begin{align}
    \alpha _{ k } ^{ } =& \frac{ N _{ k } }{ N }\\
    \label{eq:mu}
    \vec{ \mu } _{ k } =& \left( \frac{ 1 }{ N _{ k } } \right) \sum _{ i = 1 } ^{ N } w _{ i k } \vec{ x } _{ i }\\
    \label{eq:sigma}
    \Sigma _{ k } =& \left( \frac{ 1 }{ N _{ k } } \right) \sum _{ i = 1 } ^{ K } w _{ i k } \left( \vec{ x } - \vec{ \mu } _{ k } \right) ^{ T } \left( \vec{ x } - \vec{ \mu } _{ k } \right)
\end{align}
Finally, we calculate the log-likelihood of the current model for these points to determine whether the process is converged.
\begin{equation}
    \label{eq:log}
    \sum _{ i = 1 } ^{ N } \log p \left( \vec{ x } _{ i } | \Theta \right) = \sum _{ i = 1 } ^{ N } \left( \log \sum _{ k = 1 } ^{ K } \alpha _{ k } p _{ k } \left( \vec{ x } _{ i } | \theta _{ k } \right) \right)
\end{equation}

We generate 1 million random points around 5 clustering centers as the testing data and use the same initial model and convergence criteria for Spark and Blaze.
The points are stored distributedly across multiple machines.

For Blaze, we implement this algorithm with 6 MapReduce operations per iteration.
The first MapReduce calculates the probability density according to Eq.~\ref{eq:prob}.
The second MapReduce calculates the membership according to Eq.~\ref{eq:membership}.
The third MapReduce accumulates the sum of memberships for each Gaussian component $N_k$.
The next two MapReduce perform the summations in Eq.~\ref{eq:mu} and Eq.~\ref{eq:sigma}.
The last MapReduce calculates the log-likelihood according to Eq.~\ref{eq:log}.
For Spark, we use the built-in implementation from the Spark MLlib library~\cite{meng2016mllib}.

\subsubsection{Nearest 100 Neighbors}

In this task, we find the 100-nearest neighbors of a point from a huge set of other points.
This is a common procedure in data analysis and recommendation systems.
We use 200 million random points for this test.

For both Spark and Blaze, we implement this task with the \emph{top k} function of the corresponding distributed containers and provide custom comparison functions to determine the relative priority of two points based on the Euclidean-distance.

\subsection{Performance Analysis}

We test the performance of both Spark and Blaze for the above tasks on Amazon Web Services (AWS).
The time for loading data from the file system is not included in our measurements.
Spark is explicitly set to use the \lstinline{MEMORY_ONLY} mode and we choose memory-optimized instances r5.xlarge as our testing environments which have large enough memory for Spark to complete our tasks.
Each r5.xlarge has 4 logical cores, 32GB memory, and up to 10 Gbps network performance.
% It is based on Intel Xeon Platinum 8000 series (Skylake-SP) processors with a sustained all core Turbo CPU clock speed of up to 3.1 GHz.

For Spark, we use the AWS Elastic MapReduce (EMR) service version 5.20.0 , which comes with Spark 2.4.0.
Since in the default setting, Spark changes the number of executors on the fly, which may obscure the results, we set the environment variable for maximizing resource allocation to true to avoid the change.
We also manually specify the number of partitions to 100 to force the cross-executor shuffle on the entire cluster.
For Blaze, we use GCC 7.3 with -O3 optimization and MPICH 3.2. 
% We also test the effects of using Thread-Caching Malloc (TCMalloc) from Google, the results of which are indicated by Blaze TCM.
For both Spark and Blaze, we perform warmup runs before counting the timings.
Timings are converted to more meaningful results for each task.

\begin{figure}
  \begin{center}
  \includegraphics[width=\linewidth]{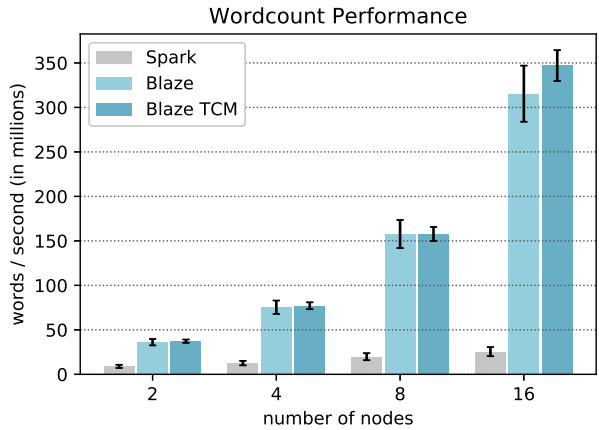}
  \end{center}
  \vspace{-0.2cm}
  \caption{Performance of the word frequency count measured in the number of words processed per second.
%   For Spark, we use the example code from its official website.
%   For Blaze, we implement the algorithm with 1 MapReduce operation.
  }
  \label{fig:wordcount_speed}
\end{figure}

\begin{figure}
  \begin{center}
  \includegraphics[width=\linewidth]{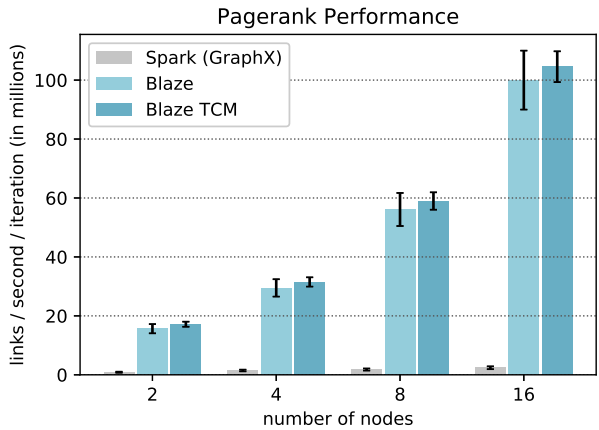}
  \end{center}
  \vspace{-0.2cm}
  \caption{Performance of the PageRank algorithm measured in number of links processed per second per iteration.
%   For Spark, we use the built-in implementation from its GraphX library.
%   For Blaze, we implement the algorithm with 2 MapReduce operations per iteration.
  }
  \label{fig:pagerank_speed}
\end{figure}

\begin{figure}
  \begin{center}
  \includegraphics[width=\linewidth]{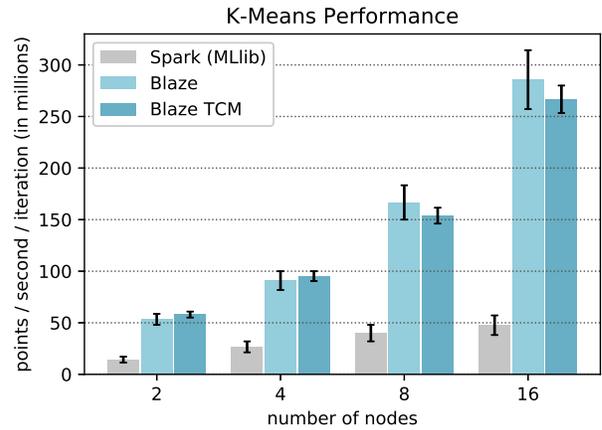}
  \end{center}
  \vspace{-0.2cm}
  \caption{Performance of the K-Means algorithm measured in the number of points processed per second per iteration.
%   For Spark, we use the built-in implementation from its MLlib library.
%   For Blaze, we implement the algorithm with 1 MapReduce operation per iteration.
  }
  \label{fig:kmeans_speed}
\end{figure}

\begin{figure}
  \begin{center}
  \includegraphics[width=\linewidth]{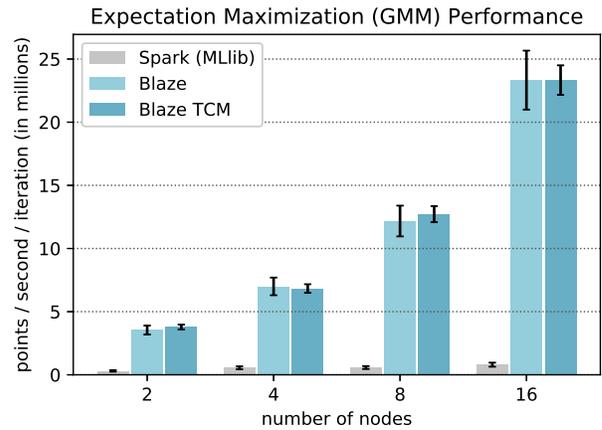}
  \end{center}
  \vspace{-0.2cm}
  \caption{Performance of the Expectation Maximization algorithm for the Gaussian Mixture Model measured in the number of points processed per second per iteration.
%   For Spark, we use the built-in implementation from its MLlib library.
%   For Blaze, we implement the algorithm with 6 MapReduce operations per iteration.
  }
  \label{fig:em}
\end{figure}

\begin{figure}
  \begin{center}
  \includegraphics[width=\linewidth]{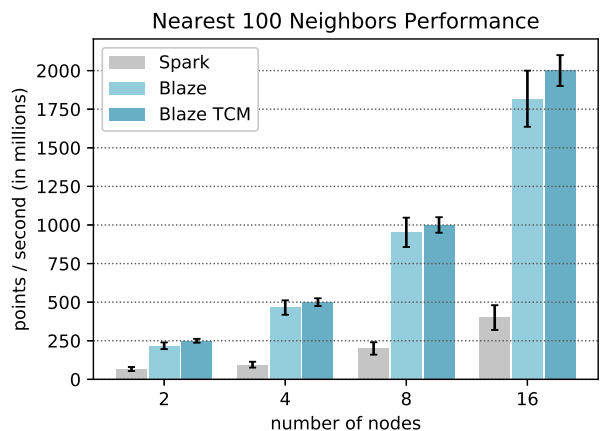}
  \end{center}
  \vspace{-0.2cm}
  \caption{Performance of the Nearest 100 Neighbors search measured in the number of total points processed per second.
%   For both Spark and Blaze, we use the top function with a custom compare function.
  }
  \label{fig:nn}
\end{figure}

% We test Spark and Blaze on the five tasks mentioned above, which are word frequency count (Worldcount), PageRank, K-Means, Expectation Maximization (EM GMM) and Nearest 100 Neighbors (NN100) Search.
The detailed performance comparison are shown in Fig.~\ref{fig:wordcount_speed} to \ref{fig:nn}.
% ``\textbf{Spark}'', ``\textbf{Spark (MLlib)}'', ``\textbf{Spark (GraphX)}'', ``\textbf{Blaze}'', ``\textbf{Blaze TCM}''
``Spark'', ``Spark (MLlib)'', ``Spark (GraphX)'', ``Blaze'', ``Blaze TCM'' denote the original Spark implementation, the MLlib library in Spark, the GraphX library in Spark, original Blaze, and Blaze linked with Thread-Caching Malloc (TCMalloc), respectively.

As shown in Fig.~\ref{fig:wordcount_speed} to \ref{fig:nn}, Blaze outperforms Spark significantly on all five data mining applications.
On average, Blaze is more than 10 times faster than Spark.
The superior performance of Blaze shows that our highly-optimized implementation suits these data mining applications well.
The performance difference between Blaze and Blaze TCM is negligible.
However, without using TCMalloc, the performance has more fluctuations and can occasionally experience a significant drop of up to 30\%.
% For specific applications, Blaze achieves at least $4 \times$ higher performance in K-Means, while it achieves $40 \times$ higher performance in PageRank.

\subsection{Memory Consumption}

\begin{figure}
  \begin{center}
  \includegraphics[width=1.0\linewidth]{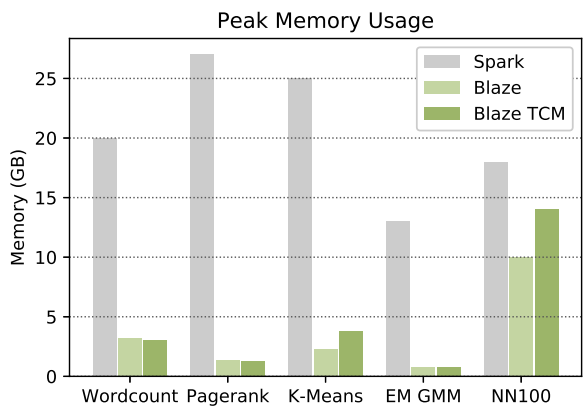}
  \end{center}
  \vspace{-0.5cm}
  \caption{Peak memory usage on a single node.
%   of 12 logical cores and 64GB memory.
%   Word frequency count task processes 350 millions words.
%   Pagerank task processes 10 million links iteratively.
%   K-means task processes 100 million points iteratively.
%   Expectation maximization (Gaussian Mixture Model) task processes 1 million points iteratively.
%   Nearest 100 neighbors task searches the nearest neighbors from 200 million points.
  }
  \label{fig:mem}
\end{figure}

We measure the memory consumption for running these tasks on a single local machine of 12 logical cores, using the same versions for all the software as the tests on AWS.
As shown in Fig~\ref{fig:mem}, we can see that both Blaze and Blaze TCM consumes much smaller amount of memory than Spark during the runs, especially for PageRank, K-Means, and expectation maximization (GMM), where Spark uses 10 times more memory than Blaze.
The only case where the memory consumption between Spark and Blaze is close is the k-nearest neighbors search, which does not involve intermediate key/value pairs.

The memory consumption between Blaze and Blaze TCM are always on the same order of magnitude, although in one case, Blaze consumes 40\% more memory when linked against TCMalloc.

\subsection{Cognitive Load}

Cognitive load refers to the efforts needed to develop or understand the code.
Minimizing the cognitive load is the ultimate goal that MapReduce and its variants try to achieve.

There are lots of different measures for cognitive efforts.
Source lines of code is not a good measure here as Spark/Scala supports chaining functions and can put several consecutive operations on a single line.
Hence, a line of Spark/Scala may be much more difficult to understand than a line of C++.
Here we use the number of distinct APIs used as the indicator for cognitive load.
It is a legitimate indicator because people will have to do more searches and remember more APIs when a library requires more distinct API calls to accomplish a task.

\begin{figure}
  \begin{center}
  \includegraphics[width=1.0\linewidth]{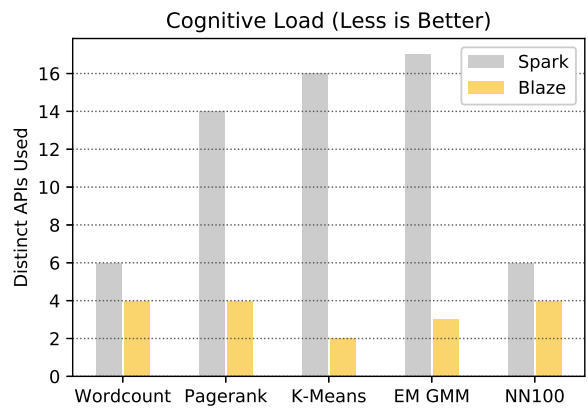}
  \end{center}
  \vspace{-0.5cm}
  \caption{Cognitive load comparison between Blaze and Spark.}
  \label{fig:cog}
\end{figure}

Spark's built-in implementation uses about 30 different parallel primitives for different tasks, while Blaze only uses the MapReduce function and less than 5 utility functions.
We can see from Fig.~\ref{fig:cog} that the cognitive load of using Blaze is much smaller than using Spark.

\section{Conclusion}
\label{sec:con}
Blaze provides a high performance implementation of MapReduce.
Users can write parallel programs with Blaze's high-level MapReduce abstraction and achieve similar performance as the hand-optimized parallel code.

% It is originally designed for compute intensive quantum simulations but we find it is also useful to many data mining tasks.
We use Blaze to implement 5 common data mining algorithms.
By writing only a few lines of serial code and apply the Blaze MapReduce function, we achieve over 10 times higher performance than Spark on these compute intensive tasks, even though we only use the MapReduce function and 3 utility functions in our Blaze implementation while Spark uses almost 30 different parallel primitives for different tasks in its official implementation.

The high-level abstraction and the high performance makes Blaze an appealing choice for compute intensive tasks in data mining and related fields.

% This paragraph will end the body of this sample document.
% Remember that you might still have Acknowledgements or
% Appendices; brief samples of these
% follow.  There is still the Bibliography to deal with; and
% we will make a disclaimer about that here: with the exception
% of the reference to the \LaTeX\ book, the citations in
% this paper are to articles which have nothing to
% do with the present subject and are used as
% examples only.
%\end{document}  % This is where a 'short' article might terminate

%ACKNOWLEDGEMENTS are optional
\section{Acknowledgements}
This work is supported by the U.S. National Science Foundation (NSF) grant ACI-1534965 and the Air Force Office of Scientific Research (AFOSR) grant FA9550-18-1-0095.
We also thank professor Cyrus Umrigar for the helpful suggestions for the paper.

% This section is optional; it is a location for you
% to acknowledge grants, funding, editing assistance and
% what have you.  In the present case, for example, the
% authors would like to thank Gerald Murray of ACM for
% his help in codifying this \textit{Author's Guide}
% and the \textbf{.cls} and \textbf{.tex} files that it describes.

%
% The next two lines define the bibliography style to be used, and the bibliography file.
\bibliographystyle{ACM-Reference-Format}
\bibliography{sample-base}

\newpage
% 
% If your work has an appendix, this is the place to put it.
% \appendix

\appendix
%Appendix A
\section{Examples}
In this section, we provide two examples to illustrate the usage of Blaze.
All the source code of our implementation is included in our GitHub repository~\cite{blaze}.

\subsection{Word frequency count}
\label{app:wordcount}
In this example, we count the number of occurrences of each unique word in an input file with Blaze MapReduce.
We save the results in a distributed hash map, which can be used for further processing.

To compile this example, you can clone our repository~\cite{blaze}, go to the \lstinline{example} folder and type \lstinline{make wordcount}.

\begin{lstlisting}
#include <blaze/blaze.h>
#include <iostream>

int main(int argc, char** argv) {
  blaze::util::init(argc, argv);
  
  // Load file into distributed container.
  auto lines =
      blaze::util::load_file("filepath...");

  // Define mapper function.
  const auto& mapper = [&](
      const size_t,  // Line id.
      const std::string& line,
      const auto& emit) {
    // Split line into words.
    std::stringstream ss(line);
    std::string word;
    while (getline(ss, word, ' ')) {
      emit(word, 1);
    }
  };

  // Define target hash map.
  blaze::DistHashMap<std::string, size_t> words;

  // Perform mapreduce.
  blaze::mapreduce<
      std::string, std::string, size_t>(
          lines, mapper, "sum", words);
    
  // Output number of unique words.
  std::cout << words.size() << std::endl;
  
  return 0;
}
\end{lstlisting}

\subsection{Monte Carlo Pi Estimation}
\label{app:pi}

In this example, we present a MapReduce implementation of the Monte Carlo $\pi$ estimation.

To compile this example, you can clone our repository~\cite{blaze}, go to the \lstinline{example} folder and type \lstinline{make pi}.

\begin{lstlisting}
#include <blaze/blaze.h>
#include <iostream>

int main(int argc, char** argv) {
  blaze::util::init(argc, argv);
  
  const size_t N_SAMPLES = 1000000;

  // Define source.
  blaze::DistRange<size_t> samples(0, N_SAMPLES);

  // Define mapper.
  const auto& mapper = 
      [&](const size_t, const auto& emit) {
    // Random function in std is not thread safe.
    double x = blaze::random::uniform();
    double y = blaze::random::uniform();
    // Map points within circle to key 0.
    if (x * x + y * y < 1) emit(0, 1);
  };

  // Define target.
  std::vector<size_t> count(1);  // {0}

  // Perform MapReduce.
  blaze::mapreduce<size_t, size_t>(
      samples, mapper, "sum", count);

  std::cout << 4.0 * count[0] / N_SAMPLES
      << std::endl;
  
  return 0;
}
\end{lstlisting}

In conventional MapReduce implementations, mapping big data onto a single key is usually slow and consumes a large amount of memory during the map phase.
Hence, in practice, people usually hand-code parallel for loops in such situations.
However, by using Blaze, the above code has similar memory consumption and performance as the hand-optimized parallel for loops.
In short, Blaze frees users from dealing with low-level data communications while ensuring high performance.

\end{document}